\documentclass[letter,superscriptaddress, prl, twocolumn]{revtex4}
\usepackage{verbatim, amsmath, graphicx}
\begin{document}
\title{ Antigravity and the big crunch/big bang transition }
\author{Itzhak Bars}
\affiliation{Department of Physics and Astronomy,
University of Southern California, Los Angeles, CA 90089-2535 USA}
\author{Shih-Hung Chen}
\affiliation{Perimeter Institute for Theoretical Physics, Waterloo, ON N2L 2Y5, Canada}
\affiliation{Department of Physics and School of Earth and Space Exploration, Arizona State University, Tempe, AZ 85287-1404 USA}
\author{Paul J. Steinhardt}
\affiliation{Department of Physics and Princeton Center for Theoretical Physics, Princeton University, Princeton, NJ 08544, USA}
\author{Neil Turok$^2$}

\begin{abstract}
We point out a new phenomenon which seems to be generic in 4d effective theories of scalar fields
coupled to Einstein gravity, when applied to cosmology. A lift of such theories to a Weyl-invariant extension 
allows one to define classical evolution through cosmological singularities unambiguously, and hence 
construct geodesically complete background spacetimes. An attractor mechanism ensures that, 
at the level of the effective theory, generic solutions undergo a big crunch/big bang 
transition by contracting to zero size, passing through a  brief antigravity phase, shrinking to zero size again, 
and re-emerging into an expanding normal gravity phase.  The result may be useful for the construction of 
complete bouncing cosmologies like the cyclic model.
\end{abstract}
\maketitle

Resolving the big bang singularity is one of the central challenges for fundamental physics and 
cosmology.  At present, there are diverse views about what form the resolution may take.  A common 
idea is that the singularity was the beginning of space and time.  In this case, 
the universe is less than 14 billion years old, and its large-scale structure must be set in place 
within the first fraction of a second.  This reasoning points to inflation \cite{inflationGuth} as the only 
rapid means of achieving the observed large-scale conditions; but then one 
is also forced to come to grips with the measure problem, the entropy problem and the fine-tuning 
problem that go hand-in-hand with inflation \cite{SciAm}.  An alternative idea is that the big bang was a bounce: a 
transition from contraction to expansion.  This idea underlies the cyclic model \cite{cyclic1}, in which the 
large-scale structure of the universe is set during an ekpyrotic contraction phase \cite{ekpyrotic1,ekpyrotic2},
well before the big bang, and then evolves through a big crunch/big bang transition.  One 
possibility for this bounce is a {\it non-singular} transition, in which the cosmic scale factor $a(t)$ rebounds at
a finite non-zero value \cite{newek}.  Einstein general relativity can describe the entire transition, but only at the price 
of introducing an energy component capable of violating the null energy condition, with the risk of undesirable 
instabilities.  Another possibility is a {\it singular} bounce \cite{cyclic1}, in which $a(t)$ shrinks to zero and 
immediately rebounds \cite{branesMT,inflationBC,cyclicBCT,BCST2,Bars15}.

In this paper, we present a novel third possibility for the bounce, involving 
a brief effective antigravity phase {\it between} the big crunch and the big bang. The
purely classical,  low-energy effective Einstein-scalar description we shall present should be 
taken only as a first indication of what may be expected when a fundamental theory 
of quantum gravity is applied to cosmological singularities. Nevertheless, we shall show that
an antigravity phase occurs generically, when we extend geodesically 
incomplete cosmological solutions to geodesically complete solutions 
of a Weyl-invariant ``lift"  of the theory: see Eq.~(\ref{conformal action}). 
(This approach for constructing  cosmological solutions \cite{inflationBC,cyclicBCT,BCST2,Bars15} is inspired 
by studies of 2T-physics \cite{2Tgravity,2TgravityGeometry}.)

A case of special significance is when the big crunch is preceded by an ekpyrotic phase, a period of ultra-slow contraction with equation-of-state $w>>1$. During an ekpyrotic phase, as the universe contracts, the homogeneous and isotropic component, represented, for example by a scalar field $\sigma$ rolling down a steep, negative potential, quickly dominates over the spatial curvature, matter density or inhomogeneities. In this way, the ekpyrotic phase smoothes the universe, resolving the cosmic horizon and flatness problems, and exponentially dilutes the anisotropies, while generating a nearly scale-invariant spectrum of density perturbations.  The ekpyrotic phase ends at a finite value of the scale factor when the scalar potential reaches a minimum. The universe continues to contract but, from this point on, the energy density is dominated by the scalar field kinetic energy, with a subdominant radiation component. (Throughout this paper, ``radiation'' refers to all forms of relativistic matter.)  Non-relativistic matter, spatial curvature and scalar potential energy (or dark energy) can be neglected the rest of the way to the crunch.  

In this Letter, we will take these
to be the initial conditions for our analysis of the big crunch/big bang transition, although more general cases (with similar results)
can be found in Ref.~\cite{BCST2}.
The effective action is that for a scalar field minimally coupled to Einstein gravity:
\begin{eqnarray}
S& = & \int d^{4}x\sqrt{-g}  [
\frac{1}{2\kappa^{2}}R(g) -\frac{1}{2}(\partial \sigma)^2 ], \label{action}%
\end{eqnarray}
where $\kappa^2\equiv 8 \pi G$, with $G$ Newton's constant. We shall also include a radiation component, parameterized by a single constant, $\rho_r$. The presence of the scalar field eliminates mixmaster chaos near the cosmic singularity and ensures that the evolution becomes {\it smoothly ultralocal}, meaning that spatial gradients become dynamically negligible \cite{Rendall}.  Although the spatial curvature and anisotropy diverge as the crunch approaches, they are both overwhelmed by the scalar field kinetic energy density.  Nevertheless, the radiation and anisotropy will each play an important role, as we shall explain.  We use the Bianchi IX metric as an illustration, discussing other cases in \cite{BCST2}. At each spatial point, the line element is:
\begin{equation}
ds^{2}  =a_{E}^{2}\left(\tau\right) \left( -d\tau^{2}+ds_{3}^{2}\right) ,\label{FRW}
\end{equation}
where the Einstein-frame scale factor $a_E(\tau)$ is a function of conformal time $\tau$ and the 3-metric $ds_3^2$ is given by
\begin{eqnarray}
e^{-\sqrt{8/3}\kappa\alpha_{1}}
d\sigma_z^{2}
+e^{\sqrt{2/3}\kappa\alpha_{1}}\left( e^{\sqrt{2}\kappa
\alpha_{2}} d\sigma_x^{2} 
+e^{-\sqrt{2}\kappa\alpha_{2}}
d\sigma_y^{2}\right)\label{Kasner}%
\end{eqnarray}
where $d\sigma_{x,y,z}$ are $SU(2)$ left-invariant one-forms,  and $\alpha_{1,2}(\tau)$ parameterize the anisotropy\cite{BCST2,Bars15}. As $a_E$ tends to zero, the dynamics simplifies. Terms involving the spatial curvature and scalar potential are suppressed by powers of $a_E$ and, provided $V(\sigma)$ is not too steep, become negligible. The four dynamical degrees of freedom, $a_E(\tau),\alpha_{1,2}(\tau)$ and $\sigma(\tau)$ obey the equations:
\begin{gather}
\frac{\dot{a}_{E}^{2}}{a_{E}^{4}}=\frac{\kappa^{2}}{3}\left[ \frac
 { \dot{\sigma}^{2} +\dot{\alpha_1}^{2} +\dot{\alpha_2}^{2} }{2a_{E}^{2}} +\frac{\rho_{r}}%
{a_{E}^{4}}\right], \label{friedmann}\\
\ddot{\sigma}+2\frac{\dot{a}_{E}}{a_{E}}\dot{\sigma} =0,\qquad 
\ddot{\alpha}_{i}+2\frac{\dot{a}_{E}}{a_{E}}\dot{\alpha}_{i} =0,\label{a1}
\end{gather}
where dot denotes $\tau$ derivative and $i=1,2$. Equations (\ref{friedmann}) and (\ref{a1}) follow from the effective action
\begin{equation}
\int d\tau\left(
\frac{1}{2e}\left[ -\frac{6}{\kappa^{2}}\dot{a}_{E}^{2}+a_{E}^{2}(\dot{\sigma}^{2}+\dot{\alpha}_{1}^{2}+\dot{\alpha}_{2}^{2})\right] 
-e \rho_{r}  \right),  \label{fieldaction}
\end{equation}
where $e(\tau)$ is the lapse function.  

The key to our approach is
to ``lift" the Einstein-scalar theory described by  (\ref{action}) to one incorporating 
Weyl-invariance. This is achieved by adding an extra scalar field and imposing 
Weyl symmetry so the new scalar degree of freedom can locally be gauged away. 
The resulting ``master" action is:
\begin{equation}
\int d^{4}x\sqrt{-g}\left[\frac{1}{2}\left((\partial \phi)^2 - (\partial s)^2\right)
+\frac{1}{12} ( \phi^{2}-s^{2})  R\right],  \label{conformal action}%
\end{equation}
to which one may add, when needed, terms representing radiation,  the scalar potential and other fields and interactions.
This Weyl-invariant action initially emerged as a $3+1$-dimensional shadow of 2T-gravity in $4+2$
dimensions \cite{inflationBC,cyclicBCT,BCST2,Bars15,2Tgravity,2TgravityGeometry}.  While the new theory is obtained from 
the Einstein-scalar theory by adding only gauge degrees of  freedom, it has an enlarged domain of field space, allowing geodesically incomplete solutions to Einstein gravity to be extended to geodesically complete solutions.    

Specifically, the master action includes two conformally coupled scalar fields and is invariant
under the local gauge transformations $g_{\mu \nu} \rightarrow \Omega^2(x^{\mu}) g_{\mu \nu}$, 
$\phi \rightarrow \Omega^{-1}(x^{\mu}) \phi$ and $s \rightarrow \Omega^{-1}(x^{\mu}) s$.  
The gravitational coupling $\kappa^2$  is replaced by $6/(\phi^2-s^2)$: for this to be positive, and the theory 
Weyl-invariant, one of the scalars, namely $\phi$, must have a wrong sign kinetic energy, potentially making it  a ghost.  However, the local Weyl gauge symmetry compensates, thus ensuring the theory is unitary. The gravitational anomaly in the trace of the stress-energy tensor cancels because $\phi$ and $s$ contribute with opposite signs~\cite{BD}.  In addition, the Lagrangian (\ref{conformal action}) possesses a global $O(1,1)$ symmetry, {\it i.e.}, the symmetry leaving $\phi^2- s^2$ unchanged. 
We do not expect this symmetry to survive quantum gravity corrections. However, it is interesting to observe that in string theory, the low-energy effective action in fact possesses a closely related, purely classical, global symmetry under shifts of the dilaton, to lowest order in the string coupling but to all orders in $\alpha'$.

A special role is played, in our analysis, by the variable 
\begin{equation}
\chi \equiv {\kappa^2\over 6} (-g)^{1\over 4} (\phi^2-s^2),
\end{equation}
which is both Weyl- and $O(1,1)$-invariant and, as we shall see, analytic at generic cosmological singularities.

We will discuss three gauge choices, in which we denote fields by the subscripts $c,E$ and $\gamma$ respectively. In the constant-gauge ($c$-gauge), we fix $\phi_c=\phi_0=const$\cite{2Tgravity}. The last term  in (\ref{conformal action})
 now takes a form similar to that found in supergravity, including the K\"{a}hler potential. The possibility that the coefficient of $R$ might switch sign was mildly noted in \cite{weinberg,deWit} but has so far been unexplored. The $c$-gauge shows that the phenomena described here, including effective antigravity, should also be expected in supergravity models and indeed, we find solutions to supergravity exhibit this behavior \cite{BCST2}.

The Einstein ($E-$gauge)  description (\ref{action}) is obtained from (\ref{conformal action}) by fixing $\frac{1}{12} (\phi_E^2 -s_E^2) = 1/2 \kappa^2>0$ , which corresponds to setting  $\phi_E = \pm (\sqrt{6}/\kappa) {\rm cosh}(\kappa \sigma/\sqrt{6})$ and  $s_E= (\sqrt{6}/\kappa) {\rm sinh}(\kappa \sigma/\sqrt{6})$. In this gauge, the vanishing of $\chi$  as $a_E\rightarrow 0$ signifies the vanishing of the determinant of the metric $g_E$, and the complete failure of the theory. Likewise, $\phi_E$ and $s_E$ typically diverge in $E$-gauge. However, in the ``lifted" theory, the problem of $g$'s vanishing may be avoided by simply fixing a different conformal gauge, for example one in which $g=-1$. In this gauge, which we denote $\gamma$-gauge,  the scale factor of the universe is unity, $a_\gamma=1$, and $\phi_\gamma$ and $s_\gamma$ remain finite in all solutions. In $\gamma$-gauge, the master action  reads
\begin{eqnarray} \label{effect}
\int d\tau\left( 
\frac{1}{2e}\left[ -\dot{\phi}_{\gamma}^2  +  \dot{s}_{\gamma}^2
+ \frac{\kappa^2}{6}(\phi_{\gamma}^2 - s_{\gamma}^2)(\dot{\alpha}_1^2+ \dot{\alpha}_2^2)\right]  -e \rho_r  \right)
\end{eqnarray}
The $E$-gauge variables $\sigma,a_{E}$ are given in terms of $\phi_{\gamma},s_{\gamma}$ as follows:
\begin{equation}
a_{E}^{2}=\left\vert \chi \right\vert,\,\chi\equiv\frac{\kappa^{2}}
{6}\left(  \phi_{\gamma}^{2}-s_{\gamma}^{2}\right),\,\sigma=\frac{\sqrt
{6}}{ 2 \kappa}\ln \left\vert \frac{\phi_{\gamma}+s_{\gamma}%
}{\phi_{\gamma}-s_{\gamma}}\right\vert \label{link1}%
\end{equation}
The cosmic singularity $a_E^2\propto \phi_{\gamma}^2-s_{\gamma}^2=0$ corresponds to the $\pm$45 degree lines in the $\phi_{\gamma}$-$s_{\gamma}$ plane, which form  the ``lightcones'' in Fig.~1.
 The singular solutions to the Friedmann equations ending in a big crunch or beginning with a big bang  correspond to trajectories confined to  $\phi_{\gamma}^2 -s_{\gamma}^2>0$ on the left and right quadrants of Fig.~1.  
The corresponding solutions for $\phi_{\gamma}$ and $s_{\gamma}$, however, can pass through all 
four quadrants \cite{cyclicBCT,BCST2,Bars15}, as shown in Fig.~1(a), including  
regions corresponding to $\phi^2 -s^2<0$, or negative Newton's constant; in other words, {\it antigravity}.  

If the anisotropy is set precisely to zero and the curvature and potential $V(\sigma)$  are non-negligible, the classical solutions typically cross the light cone in the  $\phi_{\gamma}$-$s_{\gamma}$ plane at any point, as illustrated Fig.~1(a).   For a special subset of parameters and initial values, the trajectory passes smoothly from the left quadrant, say, through the origin $\phi_{\gamma}=s_{\gamma}=0$   and onwards  to the right quadrant
\cite{cyclicBCT,BCST2,Bars15}. The universe shrinks to zero size ($a_E(\tau)=0$)  at the singularity and rebounds without encountering any region with effective antigravity. This zero-size bounce does not require any violation of the null energy condition.

\begin{figure}[t]
\includegraphics[width=2.45in]{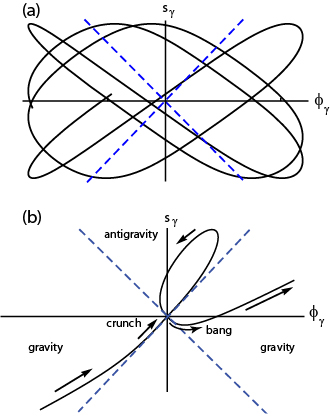}
\caption{Trajectories in the  $\phi_{\gamma}$-$s_{\gamma}$ plane for (a) typical solution with no anisotropy; (b) generic solution with anisotropy, which produces an attractor effect that draws the trajectories to the origin, $\phi_\gamma=s_\gamma=0$, at the big crunch, through a finite antigravity loop, and out of the origin again in a big bang.} 
\end{figure}

However, when anisotropy is included, no matter how small, the solutions are qualitatively different. 
 To see this, it is useful to write the equations of motion (\ref{friedmann}) and (\ref{a1})
in terms of the canonical momenta $\pi_\sigma=a_{E}^{2}\dot{\sigma},$  $\pi_1= a_{E}^{2}\dot{\alpha}_{1},$ 
$\pi_2=a_{E}^{2}\dot{\alpha}_{2}$. These are conserved when the potential $V$ and the curvature are negligible, taking constant values $(\pi_\sigma,\pi_1,\pi_2)\rightarrow (p_\sigma,p_1,p_2)$. 
Then the extension of the  Friedmann equation  Eq.~(\ref{friedmann}) 
to the full $\phi$-$s$ plane is  
\begin{equation}
\dot{\chi}^{2}=\frac{2\kappa^{2}}{3}\left(  p^2   +2 \rho_{r}\chi
\right),  \label{adot}%
\end{equation}
where  $p \equiv \sqrt{p_{\sigma}^{2}+ p_{1}^{2}+p_{2}^{2}}.$   Note that the $p^2$ term, associated with the scalar kinetic energy and anisotropy,  dominates as $\chi \rightarrow 0$.   Generic solutions, with arbitrarily small but nonvanishing anisotropy, are drawn to the origin, similar to the zero-size bounces described above. However, instead of passing directly through the origin to the right gravity region, they first undergo a finite loop in the upper (or lower) antigravity quadrant before returning to the origin and passing out to the right; see Fig. 1(b).   The special quantity $\chi$, invariant under both Weyl and $O(1,1)$ symmetries, obeys a regular equation, (\ref{adot}), and is analytic throughout.

To be precise, the solution for $\phi_{\gamma}(\tau)$ and $s_{\gamma}(\tau)$ is:
\begin{align}
{\kappa\over\sqrt{6}}( \phi_{\gamma}  +s_{\gamma})   &
=\sqrt{T}  (p+\rho_r \overline{\tau})
\left|  \frac{\overline{\tau}}{T(p+\rho_r \overline{\tau})} \right|
^{(p+p_{\sigma})/2 p}
\label{f+s}
\\
{\kappa\over\sqrt{6}}(\phi_{\gamma}  -s_{\gamma})  &
=\frac{2\overline{\tau}}{\sqrt{T}}\left|  \frac{ \overline{\tau}  }{T(p+\rho_r \overline{\tau})} \right|
^{-(p+p_{\sigma})/2 p}
\end{align}
where $\overline{\tau}=\kappa \tau/\sqrt{6}$ and
$T$ is an integration constant.
Their product gives $\chi(\tau)$ in both the gravity and antigravity portions of the trajectory,
\begin{equation}
\chi\left(  \tau\right)  =2\overline{\tau} (p+\rho_r \overline{\tau})
  ,\;\;a_{E}^{2}\left(  \tau\right)  =\left\vert \chi \left(
\tau\right)  \right\vert , \label{z}%
\end{equation}
and their ratio gives $\sigma\left(  \tau\right)  $ through Eq.(\ref{link1}),%
\begin{equation}
{\kappa\over\sqrt{6}} \sigma\left(  \tau\right)  =\frac{p_{\sigma}}{2 p}
\ln |\frac{ \overline{\tau}}{T (p+\rho_r  \overline{\tau})}|. \label{sigmasoln}
\end{equation}
For the $\alpha_i$, the solution is the same as for $\sigma$ except that $p_{\sigma}$ and $T$ are replaced by $p_i$ and $T_i$.
The solutions for $\chi$ and $\sigma$ are plotted in Fig.~2.  While $\sigma$ and the $\alpha_i$ diverge at the singularities, we can construct a complete set of quantities which are finite everywhere. Returning to the effective action (\ref{fieldaction}), we observe that near the singularities the radiation term can, to a first approximation, be neglected. Setting $\kappa \alpha_0\equiv \sqrt{3/2} \ln |\chi|$ and $\sigma\equiv \alpha_3$, the master action (\ref{effect}) becomes that for a massless relativistic particle in a conformally flat  spacetime, with line element $\chi (-d \alpha_0^2+d\alpha_1^2+d\alpha_2^2+d\alpha_3^2)\equiv \chi \eta_{\mu \nu}d\alpha^\mu d\alpha^\nu$. This action is invariant under the global conformal group $O(4,2)$. As a consequence, when $\rho_r=0$, there are 15 conserved Noether charges, including the  momenta $\pi_{\mu}=\chi \eta_{\mu \nu} \dot{\alpha}^\mu/e$, angular momenta $M_{\mu \nu}=\alpha_\mu\pi_\nu-\alpha_\nu\pi_\mu$, dilatation generator $D=\alpha^\mu \pi_\mu$ and special conformal generators. These quantities are all finite at the singularity. Continuing $\chi$ analytically through the singularity (as in (\ref{z})), and matching the $O(4,2)$ generators across it, uniquely determines the solution to be that given above. In \cite{BCST2}, we show that this solution is also selected by minimizing the action, including variations of all parameters describing the passage through the singularity. It is also the unique solution which extends to the complex $\tau$-plane. 

Note that $\chi(\tau) \propto \phi_{\gamma}^2(\tau) - s_{\gamma}^2(\tau)$  has two zeroes: one at the crunch 
$\tau=\tau_c=-\sqrt{6}p/(\kappa\rho_r) $ and one at the bang $\tau=0$. In between, the trajectory passes through a finite effective antigravity phase (see Fig.~2), during which the scalar kinetic energy and radiation densities contribute with opposite signs in the Friedmann equation (\ref{adot}). The former redshifts away more rapidly and, when  the two  are equal, the scale factor recontracts. The proper time spent in the antigravity loop is $ \int_{\tau_c}^0 a_E(\tau) d\tau =\sqrt{3} \pi p^2/( 4\kappa  \rho_r^{3\over 2})$. 

\begin{figure}
\includegraphics[width=2.65in]{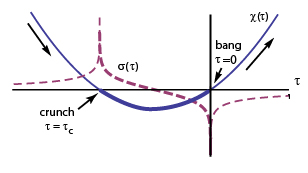}
\caption{Plots of $\chi(\tau)$ and $\sigma(\tau)$ in a big crunch/big bang transition punctuated by a brief period of antigravity between $\tau=\tau_c$ and $\tau=0$ (thick portions of the curves). }
\end{figure}

We emphasize that our results here are purely classical, based on extending the Einstein equations in the most natural way consistent with symmetries anticipated from string theory, quantum gravity and relativity to apply near a big crunch/big bang transition.  As mentioned above, our simple model is not an isolated case: in models of supergravity coupled to matter \cite{weinberg,deWit}, the effective gravitational coupling can also become negative, and we have found
analogous solutions with an antigravity phase in supergravity models~\cite{BCST2}. The antigravity phase should, we believe,  be taken as a manifestation, within the low-energy effective theory, of new physical phenomena whose detailed interpretation will require further technical developments. In this sense, it may be analogous to the Klein paradox in relativistic quantum mechanics, which correctly signaled pair production from the vacuum even before quantum field theory was developed. Although we are still far from a complete theory of quantum gravity, we may nevertheless anticipate progress in understanding the implications of an effective antigravity phase based on currently available tools. The obvious problem is that  spin-2 gravitons (as well as space-dependent fluctuations in the $\sigma$ field) have wrong-sign kinetic terms in such a phase, rendering the vacuum unstable to spontanous production of negative energy gravitons and positive energy matter particles.  However, our results suggest an interesting backreaction: particle production increases the radiation density $\rho_r$, which shortens the proper duration $\sqrt{3} \pi p^2/( 4\kappa  \rho_r^{3\over 2})$ of the antigravity loop. This  suggests a natural mechanism for cutting off the instability and at the same time producing an enhanced radiation density when the universe emerges in a big bang.  A complete picture also requires inclusion of quantum gravity effects. We have performed an analysis based on the Wheeler-de Witt equation, in the ultralocal limit, and found the same antigravity phase. To study similar phenomena in string theory, including $\alpha'$ corrections, we are investigating a Weyl-lifted version of string theory. Calculations of particle production and the evolution of classical perturbations across the bounce will also be presented elsewhere~\cite{BCST2}.

We thank Niayesh Afshordi, Tom Banks, Latham Boyle,
Bernard deWit, Jaume Gomis, Igor Klebanov and Martin Rocek for helpful conversations. Research at Perimeter Institute is supported by the Government of Canada through Industry Canada and by the Province of Ontario through the Ministry of Research and Innovation. 
This research was partially supported by the U.S. Department of Energy under grant number DE-FG03-84ER40168 (IB) and under grant number DE-FG02-91ER40671 (PJS). IB thanks Perimeter Institute for partial support during a recent sabbatical.

\end{document}